\documentclass[times,art10,twocolumn,latex8,conference]{IEEEtran}
\IEEEoverridecommandlockouts
\usepackage{cite}
\usepackage{amsmath,amssymb,amsfonts}
\usepackage{algorithmic}
\usepackage{graphicx}
\usepackage{textcomp}
\usepackage{xcolor}
\usepackage{comment}
\usepackage{breakurl}
\usepackage{xspace}
\usepackage{pifont}
\usepackage{multirow}
\usepackage{amsfonts}
\usepackage{listings}
\usepackage{subfigure}
\usepackage{xcolor}
\usepackage{color}
\usepackage{url}
\usepackage{cite}
\usepackage{booktabs}

\usepackage{tcolorbox}
\usepackage{hyperref}

\newcommand{\tool}{\texttt{MSCoT}}

\begin{document}

\title{{\tool}: Structured Chain-of-Thought Generation for Multiple Programming Languages}

\author{\IEEEauthorblockN{Naizhu Jin\IEEEauthorrefmark{2},  
Zhong Li\thanks{* Zhong Li is the corresponding author.}\IEEEauthorrefmark{2}\IEEEauthorrefmark{1},  
Tian Zhang\IEEEauthorrefmark{2},
Qingkai Zeng\IEEEauthorrefmark{2}
}

\IEEEauthorblockA{\IEEEauthorrefmark{2}\textit{State Key Laboratory for Novel Software Technology},
\textit{Nanjing University}, China\\
Email: jnz@smail.nju.edu.cn, lizhong@nju.edu.cn, ztluck@nju.edu.cn, zqk@nju.edu.cn}
}

\maketitle

\begin{abstract}
  With the rapid development of code intelligence, the application of multiple programming languages is becoming increasingly widespread. However, most existing code generation models mainly focus on a single or a few programming languages, resulting in unsatisfactory performance in a multilingual environment.
  Chain-of-Thought (CoT) reasoning can significantly improve the performance of the model without the need for retraining or fine-tuning the code generation model by reasonably decomposing complex code generation tasks into multiple subtasks and gradually deriving solutions for each subtask.
  Nevertheless, the existing CoT generation methods mainly concentrate on Python code, and the performance on other programming languages remains unclear.
  
  To fill this gap, we first constructed a CoT generation dataset for 12 programming languages through multi-agent technology. On this basis, we proposed a CoT generation method {\tool} applicable to multiple programming languages. By introducing CoT into the code generation large model, the performance of the code generation large model in a multilingual environment can be improved.
  Through large-scale empirical research, we compared the generalization abilities of {\tool} and the existing CoT generation methods on multiple programming languages and proved the effectiveness of {\tool} for multiple programming languages. In addition, we also designed a human study to prove the quality of the CoT generated by {\tool}.
  Finally, we open-sourced the model and dataset of {\tool} to promote the research on CoT generation for multiple programming languages.  
\end{abstract}


\section{Introduction}

\begin{figure}[t]
    \centering
    \includegraphics[width=0.45\textwidth]{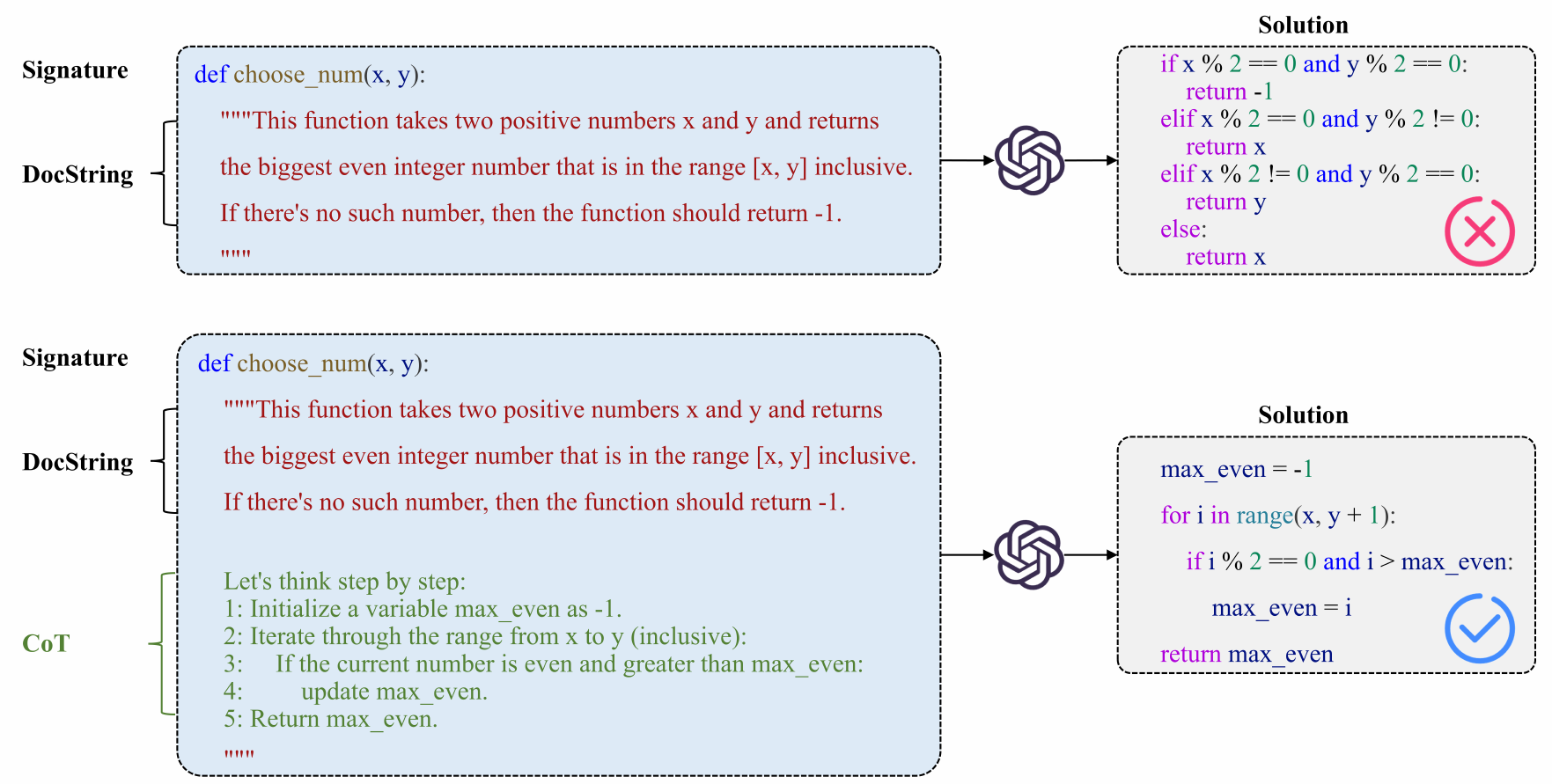}
    \caption{Example of CoT Generation}
    \label{fig:example}
\end{figure}

With the rapid development of Large Language Models (LLMs), Code Language Models (CLMs) are widely used in the software development process for code generation~\cite{vaithilingam2022expectation,yang2024important}. To enhance CLMs' effectiveness without retraining or fine-tuning, Chain-of-Thought (CoT) reasoning has emerged as a promising frontier technology~\cite{wei2022chain}. As demonstrated in Fig.~\ref{fig:example}, CoT technology analyzes the signatures and docstring input, systematically decomposes complex code generation tasks into subtasks, and methodically derives solutions for each component.

\begin{figure}[t]
    \centering
    \includegraphics[width=0.45\textwidth]{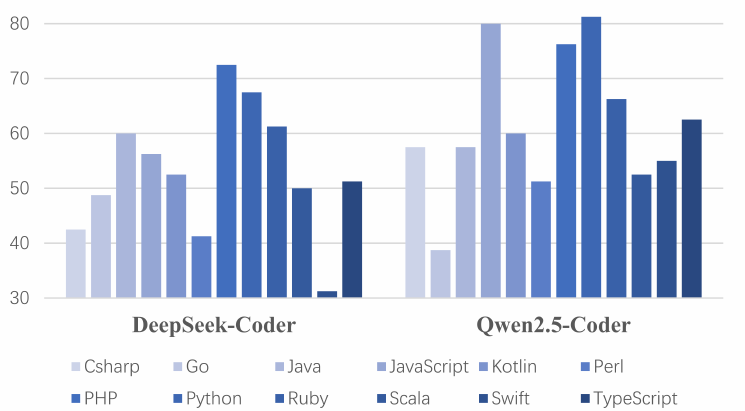}
    \caption{Performance of Code Generation Models on HumanEval-XL}
    \label{fig:motivation}
\end{figure}

Current CoT generation methods heavily depend on manual CoT writing or large-scale LLMs exceeding 100B parameters (e.g., GPT-4), resulting in substantial costs and limited applicability in resource-constrained scenarios~\cite{jiang2024self, yang2024chain}. Notably, the improved SCoT method~\cite{li2023structured} constrains large models(e.g., GPT-4) through structured chain-of-thought reasoning, organizing processes around three fundamental program structures (sequential, conditional, and iterative). More recently, Yang et al.~\cite{yang2024chain} developed CodeCoT-9k, a dataset containing 9,000 CoT samples, and proposed a lightweight CoT model named COTTON for CoT generation. However, their approach primarily focuses on Python, overlooking the diverse syntax of signatures and the style of docstrings across different programming languages.

The primary challenge in training CoT models lies in the scarcity of high-quality training data, which requires domain expertise and is time-consuming to construct. The challenge of data construction is further amplified in multilingual scenarios, where variations in syntax and documentation styles complicate the creation of high-quality CoT datasets. As software development grows more complex and diverse, multi-language integration has become an irreversible trend~\cite{gao2024current, uchitel2024scoping}. For instance, in web development, JavaScript is the cornerstone for dynamic web applications through its asynchronous capabilities and rich front-end framework ecosystem (e.g., React, Vue.js, Angular). Meanwhile, backend development typically relies on Java, Python, or Go. In data science and AI, Python dominates due to its comprehensive data processing libraries, machine learning frameworks, and scientific computing tools (such as TensorFlow and PyTorch). In this sophisticated multi-language development environment, the importance of code generation technology capable of effectively supporting multiple programming languages has become increasingly prominent~\cite{zheng2023survey, cassano2024knowledge, yang2024multi}. 
However, existing Code LLMs predominantly focus on single or limited programming languages, resulting in suboptimal performance in multi-language environments.
Consider DeepSeek-Coder 6.7b-instruct and Qwen2.5Coder 7b-instruct, both trained on multi-language datasets. 
Our evaluation on HumanEval-XL~\cite{peng2024humaneval} reveals significant performance disparities across different programming languages, as illustrated in Fig.~\ref{fig:motivation}.

Addressing these challenges, we introduce a large-scale CoT dataset encompassing 12 programming languages, developed through multi-agent technology. 
Our multi-agent interaction framework comprises three specialized agents: Code Quality Agent (CQAgent), Code Translation Agent (CTAgent), and Structured Chain-of-Thought Agent (SCoTAgent).
CQAgent evaluates code samples based on educational value and docstring-implementation consistency. 
CTAgent performs style-aware code translation across languages using few-shot learning while maintaining semantic equivalence. 
SCoTAgent generates structured chain-of-thought explanations by given the docstring and signature through the three fundamental program structures.
Finally, we have constructed 7,000 CoT samples for each of the 12 programming languages(i.e., CSharp, Go, Java, JavaScript, Kotlin, Perl, PHP, Python, Ruby, Scala, Swift, and TypeScript), totaling 84,000 samples.
Furthermore, to optimize CoT generation for resource-constrained scenarios, we fine-tuned the 7B-parameter Qwen2.5Coder model using LoRA technique to develop the final {\tool} model.

Beyond dataset construction, we introduce a novel Instruction Template designed to guide LLMs in generating more structured and informative CoT explanations. This template standardizes the reasoning process and improves model interpretability, making CoT generation more effective across diverse programming tasks.

To validate {\tool}'s quality and generalization capabilities, we evaluated its performance using two state-of-the-art CodeLLMs(i.e., DeepSeek-Coder 6.7b-instruct and Qwen2.5Coder 7b-instruct) across 12 programming language benchmarks.
Compared to zero-shot and other baseline CoT generation methods, {\tool} demonstrates significant improvements in multi-language CoT generation while maintaining cost-effectiveness.
Additionally, we provide comprehensive analysis explaining the performance instability of baseline methods.
Finally, we conduct human evaluation studies to validate the quality of {\tool}-generated CoTs.

In summary, the main contributions of our work can be summarized as follows:
\begin{itemize}
  \item We construct a large-scale CoT dataset contains 84,000 samples covering 12 programming languages through a multi-agent framework, and then propose {\tool} which supports multi-language CoT generation. 

  \item We conduct comprehensive experiments across multiple programming languages and state-of-the-art CodeLLMs to validate {\tool}'s effectiveness. 
  
  \item We open-source our model\footnote{\url{https://modelscope.cn/models/cotmodel/MSCoT}} and dataset\footnote{\url{https://github.com/WolfgangJin/MSCoT}} to facilitate future research in multi-language CoT generation and code intelligence.
\end {itemize}

\section{Background and Related Work}

\subsection{Code Generation}
Code generation is a key task in automated software development, focusing on transforming functional descriptions into executable code. Formally, the task can be described by a dataset $\mathcal{D} = \{(X_i, Y_i)\}_{i=1}^{|\mathcal{D}|}$, where $X_i$ represents a functional description and $Y_i$ denotes the corresponding code snippet. The objective of a code generation model $M_{code}$ is to predict $Y_i$ based on $X_i$. This can be formulated as an autoregressive process, parameterized by $\theta_{code}$:
\[
P_{\theta_{code}}(Y_i \vert X_i) = \prod_{k=1}^{n} P_{\theta_{code}}(Y_{i,k} \vert X_i, Y_{i,1}:Y_{i,k-1}),
\]
where $Y_{i,1}:Y_{i,k-1}$ denotes the sequence of tokens generated before the $k$-th token, and $n$ is the total number of tokens in $Y_i$.

Early research on neural code generation often relied on heuristic rules and expert systems (e.g., those based on probabilistic grammars~\cite{cohn2010inducing, allamanis2014mining} or domain-specific languages~\cite{gulwani2010dimensions, zan2023large}), but such approaches generally lacked flexibility and scalability~\cite{zan2023large}. As technology advanced, attention shifted to deep learning models such as CNNs~\cite{liu2016automatic, sun2019grammar}, RNNs~\cite{iyer2016summarizing, wan2018improving}, and LSTMs~\cite{yin2017syntactic}, as well as the Transformer architecture~\cite{vaswani2017attention}, originally developed for machine translation in 2017 and quickly adopted for code generation~\cite{mastropaolo2021studying, shah2021natural}. However, these models typically require large volumes of labeled data for training, and their performance remains somewhat limited.

With the advent of pre-training and fine-tuning techniques, models featuring fewer than 1 billion parameters—such as CodeBERT~\cite{feng2020codebert}, CodeGPT~\cite{lu1codexglue}, PLBART~\cite{ahmad2021unified}, and CodeT5~\cite{wang2021codet5}—have shown promising results in multilingual code generation. More recently, large language models with over 10 billion parameters have demonstrated zero-shot code generation capabilities. Among these, Codex (12B parameters)~\cite{chen2021evaluating} stood out in Python coding tests, ultimately leading to its integration into commercial products like Copilot\footnote{\url{https://github.com/features/copilot}}. Following Codex, specialized LLMs such as AlphaCode~\cite{li2022competition}, InCoder~\cite{friedincoder}, CodeGen~\cite{nijkampcodegen, nijkampcodegen2}, StarCoder~\cite{li2023starcoder}, WizardCoder~\cite{luowizardcoder}, OctoCoder~\cite{muennighoff2023octopack}, and CodeLlama~\cite{roziere2023code} have emerged, revealing potential to tackle complex programming problems and support developers in various environments.

\begin{figure*}[t]
  \centering
  \includegraphics[width=0.8\textwidth]{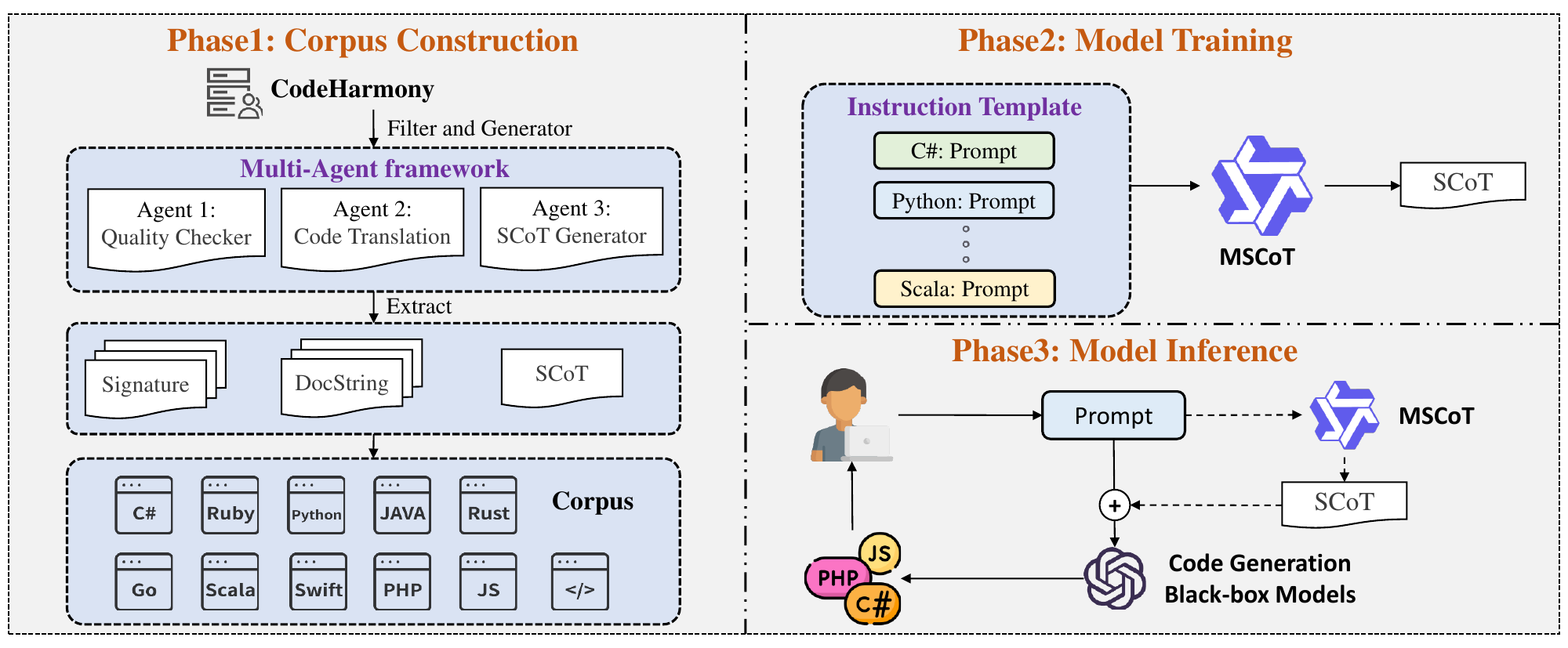}
  \caption{Approach of {\tool}}
  \label{fig:approach}
\end{figure*}

\subsection{Chain-of-Thought Generation}
The CoT generation model $M_{cot}$ aims to produce a well-formed CoT $C_i$ based on $X_i$. 
This CoT is then appended to the original input sequence $X_i$, forming a new sequence $\hat{X}_i = X_i \oplus C_i$, where $\oplus$ denotes concatenation. 
The probability of generating the code snippet $Y_i$ conditioned on $X_i$ is approximated as:
\[
P(Y_i \vert X_i) \propto \underbrace{P_{\theta_{cot}}(C_i \vert X_i)}_{M_{cot}} \underbrace{P_{\theta_{code}}(Y_i \vert X_i, C_i)}_{M_{code}}.
\]

As the scale of model parameters and the amount of training data continue to expand, LLMs have achieved significant progress in their reasoning abilities~\cite{wei2022emergent}. Chain-of-Thought (CoT) prompting has emerged as a key technique for improving performance by systematically breaking down problems and providing detailed explanations~\cite{wei2022chain}.

In the context of code generation, researchers have begun exploring CoT's potential. 
Li et al.~\cite{li2023structured} introduced a structured CoT approach to help models better understand complex intentions. 
Similarly, Jiang et al.~\cite{jiang2024self} proposed a self-planning method, aiming to improve the model’s ability to plan and reason autonomously during code generation. 

Despite progress, deploying LLMs with over 100 billion parameters remains challenging due to high computational demands. Researchers address this by using smaller models and techniques like knowledge distillation to achieve complex reasoning efficiently.
For instance, Yang et al.~\cite{yang2024chain} proposed COTTON, which constructs a CoT dataset for code generation using multi-agent technology and leverages lightweight models for learning CoT generation.

\section{Method}
The workflow of our proposed {\tool} is shown in Fig.~\ref{fig:approach}. 

\subsection{Dataset Construction}
Generating training data directly with Large Language Models (LLMs) often leads to hallucinations, making it unreliable for constructing a high-quality dataset~\cite{zhang2024self}.
To mitigate these issues, we require a high-quality dataset as a seed for expansion which should support function-level code generation and include rich documentation to facilitate effective CoT generation. Based on these criteria, we selected CodeHarmony~\cite{yang2024less} as our raw datasets, which focus on function-level code generation and extract well-established open-source repositories, primarily the Evol dataset~\cite{luowizardcoder} and the OSS dataset~\cite{wei2024magicoder}. 
Then we design a multi-agent framework consisting of three specialized agents: CQAgent, CTAgent, and SCoTAgent. Each agent is responsible for a specific task in the dataset construction pipeline, as illustrated in Fig.~\ref{fig:approach}.

\subsubsection{Code Quality Agent (CQAgent)}
To ensure high-quality code samples, CQAgent employs a two-stage filtering process. 
First, it evaluates whether code samples possess educational value by examining their alignment with textbook standards and their effectiveness in teaching programming concepts. 
Second, it verifies the docstring-implementation consistency by attempting to generate code based on the given docstring and checking if the generated implementation matches the original code. 
Only samples that pass both quality checks are retained in the dataset.

\begin{tcolorbox}[
  width=1.0\linewidth, 
  title={CQAgent},
  colback=lightgray!20, 
  colframe=black!80, 
  boxrule=1pt, 
  arc=3mm, 
  fonttitle=\bfseries 
]
  \textbf{\textcolor{blue}{Role:}} You are a helpful code assistant. \\
  \textbf{\textcolor{blue}{Task:}} Your task is to check if the given code has the education value and whether its quality meets textbook standards, and also check if you can generate a candidate code that matches the given code based on the docstring in the code. \\
  \textbf{\textcolor{blue}{Return:}} Output True if both conditions are met, otherwise output False.
  \end{tcolorbox}

\subsubsection{Code Translation Agent (CTAgent)}
CTAgent performs style-aware code translation across different programming languages while maintaining semantic equivalence. The translation process consists of two key components: Signature Translation and Docstring Translation.

Specifically, \textbf{Signature Translation} translates function signatures between programming languages while preserving the original function name, parameter types, and return types. This process accounts for language-specific syntax and conventions, such as type declarations and function definition styles.
\textbf{Docstring Adaptation} adapts docstrings to follow the target language's documentation conventions (e.g., converting Python's style to TypeScript's style) while preserving the semantic content. 

To ensure code translation quality, CTAgent employs few-shot learning with carefully selected examples that demonstrate proper style and convention mapping between language pairs. The agent validates translations by checking both syntactic correctness and semantic preservation.

\begin{tcolorbox}[
  width=1.0\linewidth, 
  title={CTAgent},
  colback=lightgray!20, 
  colframe=black!80, 
  boxrule=1pt, 
  arc=3mm, 
  fonttitle=\bfseries 
]
  \textbf{\textcolor{blue}{Role:}} You are a helpful code assistant.\\
  \textbf{\textcolor{blue}{Task:}} Your task is to translate the following docstring and signature from \textcolor{red}{Source Code} to \textcolor{red}{Target Code}, which needs follow the format of given example code. \\
  $[$\textit{Signature Translation}$]$ translates function signatures between programming languages. \\
  $[$\textit{Docstring Translation}$]$ adapts docstrings to follow the target language's documentation conventions. \\
  \textbf{\textcolor{blue}{Return:}} Only output the docstring and signature, no other information. \\
  \textbf{\textcolor{blue}{Example Input:}} \\
  def below\_zero(operations) -\textgreater bool:\\
  ``` You're given a list of (\textit{more information})\\
  '''\\
  \textbf{\textcolor{blue}{Example Output:}} \\
  /**\\
  * You're an expert TypeScript programmer\\
  * You're given a list of (more information) \\
  */\\
  const below\_zero = function (operations): boolean \{\\
  \textbf{\textcolor{blue}{Input:}} $[X]$\\
  \textbf{\textcolor{blue}{Output:}} $[Y]$
\end{tcolorbox}

\subsubsection{Structured Chain-of-Thought Agent (SCoTAgent)}
Following the SCoT framework~\cite{li2023structured}, SCoTAgent generates the structured CoT, which follows the three fundamental program structures (sequential, conditional, and iterative) to ensure consistency and comprehensibility across different programming languages.

Specifically, SCoTAgent first generates structured CoT for Python code samples. 
Based on the observation that programming logic remains consistent across languages despite syntactic differences, we leverage the language-agnostic nature of SCoT reasoning. 
This allows us to maintain a one-to-many relationship between CoT and their corresponding implementations across different programming languages, significantly improving the efficiency of our dataset construction process while ensuring reasoning consistency.

Through the collaborative effort of these three agents, we successfully constructed a dataset containing 7,000 CoT samples for each of the 12 programming languages, totaling 84,000 high-quality samples. The dataset covers a wide range of programming tasks and ensures consistency in both code quality and CoT across different programming languages.
\begin{tcolorbox}[
  width=1.0\linewidth, 
  title={SCoTAgent},
  colback=lightgray!20, 
  colframe=black!80, 
  boxrule=1pt, 
  arc=3mm, 
  fonttitle=\bfseries 
]
  \textbf{\textcolor{blue}{Role:}} You are a helpful code assistant.\\
  \textbf{\textcolor{blue}{Task:}} Please understand the requirement and write a rough solving process. \\
  It starts with Let's think step by step and then a input-output structure. \\
  You should use three basic structures to build the solving process, including sequences, branches, and loops. \\
  The necessary details should writen in nature language.\\
  \textbf{\textcolor{blue}{Return:}} Only output the solving process, no other information. \\
  \textbf{\textcolor{blue}{Example Input:}} $[$Demo\_Input$]$ \\
  \textbf{\textcolor{blue}{Example Output:}} $[$Demo\_Output$]$ \\
  \textbf{\textcolor{blue}{Input:}} $[X]$\\
  \textbf{\textcolor{blue}{Output:}} $[Y]$
\end{tcolorbox}

\subsection{Model Training}
To effectively fine-tune {\tool} for multi-language CoT generation, we first introduce an Instruction Template that provides structured prompts to enhance reasoning consistency across programming languages. This template standardizes task-specific instructions, guiding the model in generating structured and interpretable CoT explanations.

\begin{tcolorbox}[
  width=1.0\linewidth, 
  title={Instruction Template},
  colback=lightgray!20, 
  colframe=black!80, 
  boxrule=1pt, 
  arc=3mm, 
  fonttitle=\bfseries 
]
\textbf{\textcolor{blue}{Role:}} You are a helpful \textcolor{red}{Language} code assistant.\\
\textbf{\textcolor{blue}{Task:}} Please understand the requirement and write a rough solving process. \\
\textbf{\textcolor{blue}{Input:}} $[prompt]$\\
\textbf{\textcolor{blue}{Output:}} $[CoT]$
\end{tcolorbox}

Second, we implement Low-Rank Adaptation (LoRA) to enable efficient model tuning. 
Let $\boldsymbol{W_{0}} \in \mathbb{R}^{d \times k}$ represent the original weight matrix. We decompose the weight updates into two low-rank matrices:
\begin{displaymath}
\Delta \boldsymbol{W} = \boldsymbol{B}\boldsymbol{A}
\end{displaymath}
where $\boldsymbol{B} \in \mathbb{R}^{d \times r}$ and $\boldsymbol{A} \in \mathbb{R}^{r \times k}$ with rank $r \ll \min (d, k)$.

The forward computation for input $\boldsymbol{X}$ becomes:
\begin{displaymath}
\bar{\boldsymbol{H}} = \boldsymbol{H} + \boldsymbol{B}\boldsymbol{A} \boldsymbol{X}
\end{displaymath}
where $\boldsymbol{H} = \boldsymbol{W_{0}} \boldsymbol{X}$ represents the original transformation.

To ensure stable training initialization, we employ asymmetric initialization: $\boldsymbol{A}$ uses Gaussian initialization while $\boldsymbol{B}$ starts from zero, guaranteeing $\Delta \boldsymbol{W} = \mathbf{0}$ at training onset. We extend this adaptation scheme across all linear transformations in the network to maximize model capacity while maintaining efficiency.

\subsection{Model Inference}
{\tool} employs greedy search during inference to ensure generation stability, selecting the highest-probability token at each step for consistent CoT output. Based on user input and the target programming language, {\tool} generates language-specific CoT reasoning steps. 
When integrated into code generation pipelines, {\tool} operates as a complementary component.

The generated CoT is only utilized when the primary code generation model fails to produce correct results. This black-box integration approach allows seamless enhancement of existing code generation systems while maintaining computational efficiency.

\section{Experiments}
To evaluate the effectiveness and benefits of our proposed approach, we mainly design the following three research questions (RQs):

\noindent\textbf{RQ1: Can {\tool} improve the performance of code generation models in multi-language environments?}

In this RQ, we want to evaluate whether {\tool} can improve the performance of code generation models compared with the other CoT generation methods in multi-language environments.

\noindent\textbf{RQ2: Can {\tool} generate high-quality CoT via human study?}

In this RQ, we want to evaluate whether {\tool} can generate high-quality CoT from the aspect of the developer's perspective.

\subsection{Code Generation Models}
Based on the performance evaluations on HumanEval leaderboard, we select the following state-of-the-art code generation models as our baselines:
\begin{itemize}
  \item DeepSeek-Coder (6.7B-instruct)~\cite{guo2024deepseek}: A specialized DeepSeek's code generation model trained on high-quality programming data across multiple languages. It demonstrates strong performance in understanding and generating code across various programming tasks.
  
  \item Qwen2.5-Coder (7B-instruct)~\cite{hui2024qwen2}: The latest version of Qwen's code-specialized model, featuring enhanced instruction-following capabilities and multi-language support. It shows competitive performance in code generation and understanding tasks.
\end{itemize}

Both models represent recent advances in code generation technology, featuring similar model sizes (around 7B parameters) and instruction-tuning capabilities, making them ideal candidates for our comparative study.

\subsection{Evaluation Metrics}
To evaluate the performance of the CoTs generated by {\tool}, we mainly focus on the Pass@1 metric and CoT-Pass@1 metric~\cite{yang2024chain}, when the CoT is used in the code generation pipeline.

\noindent \textbf{Pass@1.} 
The Pass@1 metric measures the percentage of generated code snippets that pass the corresponding test cases in a zero-shot setting (without CoT guidance).

\noindent \textbf{CoT-Pass@1.} 
The CoT-Pass@1 metric measures the percentage of successful code generations when CoT guidance is provided after an initial failed attempt. This metric evaluates the effectiveness of CoT in improving code generation performance.

\subsection{Baselines CoT Generation Methods}
We select the following widely used CoT generation methods as our baselines, where the first two methods rely on the code generation model itself, and the latter two methods rely on an external model.
\begin{itemize}
  \item Zero-Shot CoT: Uses the simple prompt "Let's think step by step" to guide model's reasoning process.
  
  \item Self-CoT: Provides few-shot CoT examples to guide the model's reasoning generation.
  
  \item SCoT(GPT4): Leverages GPT-4 to generate SCoT reasoning based on example patterns.
  
  \item COTTON: Employs fine-tuned CodeLlama-7B to generate CoT reasoning for code generation tasks.
\end{itemize}

\begin{table}[t]
    \centering
    \caption{Hyper-parameters and their values}
    \begin{tabular}{c|c||c|c}
    \toprule
       Hyper-parameter  & Value &  Hyper-parameter  & Value\\
     \midrule
      Optimizer & AdamW & Random Seed & 42 \\
      Learning Rate  & 2e-4 & Training batch size & 1 \\
      Lora R & 32 & Lora alpha & 16 \\
      Max input length & 512 & Max output length & 512 \\
      \bottomrule
    \end{tabular}
    \label{Hyper-parameters}
\end{table}

\subsection{Implementation Details and Running Platform}
The hyper-parameters are tuned according to actual performance. We show the values of these hyper-parameters in Table~\ref{Hyper-parameters}. 
For the implementation of {\tool} and other baselines, we utilize the PyTorch \footnote{\url{https://pytorch.org/}} and Transformers \footnote{\url{https://github.com/huggingface/transformers}} libraries.
All experiments are conducted on a machine running Ubuntu 18.04 with an Intel(R) Xeon(R) Platinum 8352V CPU, 90GB RAM, and a GeForce RTX 4090 GPU (24GB RAM). Model training for {\tool} took approximately 12 hours.

\section{RESULT ANALYSIS}

\begin{table*}[t]
  \caption{Multi-language Code Generation Performance with the guidance of {\tool} and other baselines}
  \begin{center}
   \vspace{-1mm}
  \setlength{\tabcolsep}{1mm}{
 \begin{tabular}{cc|ccccccccccccc}
   \toprule
 Model & Method & CSharp & Go & Java & JavaScript & Kotlin & Perl & PHP & Python & Ruby & Scala & Swift & TypeScript & Avg.\\
 \midrule 
 \multirow{6}{*}{\textbf{DeepSeek-Coder}} 
  & Zero-Shot & 42.50 & 48.75 & 60.00 & 56.25 & 52.50 & 41.25 & 72.50 & 67.50 & 61.25 & 50.00 & 31.25 & 51.25 & 52.92\\
  & Zero-Shot CoT & \underline{\textit{46.25}} & 55.00 & 67.50 & 62.50 & 58.75 & 46.25 & 75.00 & 71.25 & 67.50 & 53.75 & 38.75 & 53.75 & 58.02 (+5.10\%)\\
  & Self-CoT & 42.50 & 50.00 & 60.00 & 56.25 & 52.50 & 43.75 & \textbf{86.25} & 68.75 & 63.75 & 55.00 & 42.50 & 61.25 & 56.88 (+3.96\%)\\
  & COTTON & \textbf{50.00} & \underline{\textit{57.50}} & 68.75 & 65.00 & 61.25 & 48.75 & 77.50 & 77.50 & 73.75 & \underline{\textit{61.25}} & 37.50 & 62.50 & 61.77 (+8.85\%)\\
  & SCoT(GPT4) & 42.50 & \textbf{62.50} & \textbf{77.50} & \textbf{75.00} & \underline{\textit{71.25}} & 57.50 & \underline{\textit{83.75}} & \textbf{90.00} & \textbf{81.25} & \textbf{62.50} & \textbf{50.00} & \textbf{77.50} & \textbf{69.27} (+16.35\%)\\
  & {\tool} & 43.75 & 60.00 & \underline{\textit{73.75}} & \underline{\textit{68.75}} & \textbf{72.50} & \textbf{60.00} & 80.00 & \underline{\textit{83.75}} & \underline{\textit{80.00}} & 60.00 & \underline{\textit{42.50}} & \underline{\textit{67.50}} & \underline{\textit{66.04}} (+13.12\%)\\
 \midrule 
 \multirow{6}{*}{\textbf{Qwen2.5-Coder}} 
  & Zero-Shot & 57.50 & 38.75 & 57.50 & 80.00 & 60.00 & 51.25 & 76.25 & 81.25 & 66.25 & 52.50 & 55.00 & 62.50 & 61.56\\
  & Zero-Shot CoT & 66.25 & 45.00 & 77.50 & 81.25 & 62.50 & 57.50 & 78.75 & 83.75 & 73.75 & 60.00 & 57.50 & 62.50 & 67.19 (+5.63\%)\\
  & Self-CoT & 57.50 & 38.75 & 58.75 & 80.00 & 61.25 & 56.25 & \textbf{100.00} & 81.25 & 67.50 & 52.50 & 57.50 & 71.25 & 65.21 (+3.65\%)\\
  & COTTON & \underline{\textit{67.50}} & \underline{\textit{47.50}} & 71.25 & \underline{\textit{82.50}} & \underline{\textit{68.75}} & 56.25 & 81.25 & 81.25 & 77.50 & 62.50 & 57.50 & 70.00 & 68.65 (+7.09\%)\\
  & SCoT(GPT4) & \textbf{73.75} & \textbf{56.25} & \textbf{80.00} & \textbf{83.75} & \textbf{73.75} & \textbf{61.25} & \underline{\textit{88.75}} & \textbf{85.00} & \textbf{82.50} & \textbf{67.50} & \underline{\textit{61.25}} & \textbf{78.75} & \textbf{74.37} (+12.81\%)\\
  & {\tool} & 66.25 & \textbf{56.25} & \underline{\textit{78.75}} & \textbf{83.75} & \underline{\textit{68.75}} & \underline{\textit{65.00}} & 86.25 & \underline{\textit{82.50}} & \underline{\textit{78.75}} & \underline{\textit{63.75}} & \textbf{62.50} & \underline{\textit{75.00}} & \underline{\textit{72.29}} (+10.73\%)\\
   \bottomrule
 \end{tabular}
 }
  \end{center}
  \label{tab:RQ1}
 \end{table*}

\begin{figure*}[h]
    \centering
    \includegraphics[width=0.8\textwidth]{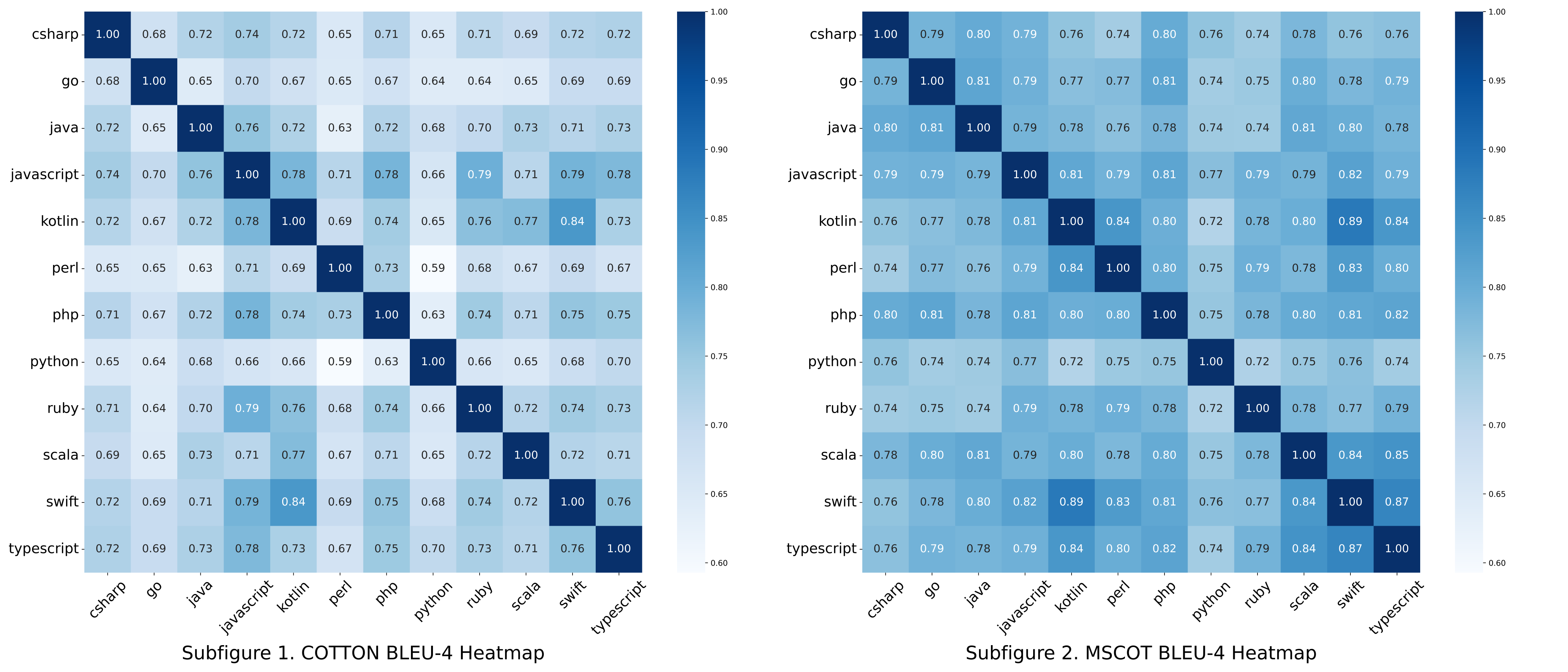}
    \caption{The correlation heatmap of the generated CoT between COTTON and MSCoT under different programming languages}
    \label{fig:heatmap}
\end{figure*}

\subsection{RQ1: Can {\tool} improve the performance of code generation models in multi-language environments?}

To answer this research question, we evaluate the performance of code generation models with and without {\tool} across 12 programming languages. The results in Table~\ref{tab:RQ1} show that integrating {\tool} with DeepSeek-Coder significantly improves the average pass@1 score from 52.92\% to 66.04\%, representing a 13.12\% absolute improvement. 
This improvement is consistent across most programming languages, with particularly notable gains in Python (+16.25\%), JavaScript (+12.50\%), and Java (+13.75\%).
When compared to existing CoT generation methods, {\tool} demonstrates superior performance. It outperforms Zero-Shot CoT by 8.02\%, Self-CoT by 9.16\%, and COTTON by 4.27\%. Most notably, {\tool}'s performance (66.04\%) approaches that of SCoT(GPT4) (69.27\%), with only a 3.23\% gap.

Similar improvements are observed with Qwen2.5-Coder, where {\tool} integration increases the average pass@1 score from 61.56\% to 72.29\%, achieving a 10.73\% absolute improvement. The enhancement is particularly evident in languages like C++ (+9.20\%), PHP (+10.00\%), and Ruby (+12.50\%), demonstrating the tool's effectiveness across diverse programming paradigms.
When compared to existing CoT generation methods, {\tool} demonstrates superior performance. It outperforms Zero-Shot CoT by 5.10\%, Self-CoT by 7.08\%, and COTTON by 4.24\%. Most notably, {\tool}'s performance (72.29\%) approaches that of SCoT(GPT4) (74.37\%), with only a 2.08\% gap. 

This close performance to GPT4-based solutions suggests that {\tool} can serve as a viable alternative in resource-constrained scenarios, offering competitive performance without requiring access to large-scale language models.

To further investigate the performance gap between COTTON and {\tool} in multilingual environments, we conducted a detailed analysis using heatmaps in Fig.~\ref{fig:heatmap} to visualize the similarity of generated CoT across different programming languages. 
In theory, CoT for the same programming task should maintain high similarity across different languages, as the underlying logical reasoning remains consistent regardless of the implementation language. 
However, our analysis reveals that COTTON's lower performance can be attributed to its insufficient adaptability across languages. 
The heatmap visualization demonstrates that COTTON's CoT generation exhibits significant variations across different programming languages, indicating its limited ability to maintain consistent reasoning patterns in a multilingual context. 

\begin{tcolorbox}[width=1.0\linewidth, title={Summary of RQ1}]
These results and the analysis suggest that {\tool}'s language-agnostic CoT generation capability effectively enhances code generation across different programming languages. 
\end{tcolorbox}

\subsection{RQ2: Can {\tool} generate high-quality CoT via human study?}

To evaluate the quality of CoTs generated by {\tool}, we conducted a human study. The study aimed to determine if they meet high standards, help in understanding code logic, and provide useful guidance across programming languages. Inspired by prior research~\cite{yang2024chain}, the evaluation focused on three aspects:
\begin{itemize}
    \item \textbf{Similarity:} This aspect measures the semantic similarity between the generated CoTs and the reference solutions.
    \item \textbf{Naturalness:} This aspect assesses the grammatical correctness and fluency of the generated CoTs.
    \item \textbf{Educational Value:} This aspect measures the ability of the generated CoTs to inspire volunteers in solving problems, reflecting their overall educational significance.
\end{itemize}

For the human evaluation, we invited one developer with over two years of programming experience for each language to participate in the study. We selected all 80 samples from HumanEval-XL. To ensure fairness, we chose five different samples for each language. For each sample, we gathered the ground truth CoT along with two generated CoTs.

\begin{table}[h]
\caption{The results of Human Study for {\tool} and COTTON}
\centering
\begin{tabular}{ccc}
\toprule
 \textbf{Aspect} & \textbf{COTTON} & \textbf{{\tool}} \\
\midrule
 Similarity &2.78 &3.47 \\
 Naturalness &2.57 &3.33 \\
 Educational Value &2.50 &3.28 \\
\bottomrule
\end{tabular}
\label{tab:RQ2}
\end{table}

The results of the human study are summarized in table~\ref{tab:RQ2}, which shows the average scores of all evaluated samples by the assessors. We can observe that the proposed approach {\tool} outperforms COTTON in terms of similarity, naturalness, and educational value. 
Building on these findings, the results for {\tool} demonstrate even greater potential in generating high-quality CoTs that surpass COTTON in all three evaluated aspects: similarity, naturalness, and educational value. The higher scores achieved by {\tool} indicate its ability to produce CoTs that are semantically closer to the ground truth and more readable and impactful in aiding developers. These improvements emphasize the robustness and adaptability of {\tool}, making it a promising solution for advancing the state-of-the-art in code generation and educational support through CoTs.

\begin{tcolorbox}[width=1.0\linewidth, title={Summary of RQ2}]
 The results show that {\tool} can generate high-quality CoTs that are semantically accurate, grammatically correct, and educationally valuable, effectively addressing the requirements of multilingual programming scenarios.
\end{tcolorbox}

\section{Threats to Validity}
The threats to external validity concern the generalizability of our findings. To address these concerns, our evaluation focused on 12 programming languages and specific code generation tasks. 
While the results may not generalize to other programming languages or more complex programming scenarios, 
the performance of {\tool} is still competitive with other CoT generation methods.

The threats to internal validity of this study are mainly from the quality and diversity of our training data. 
To avoid the risk of internal validity, we have made efforts to ensure data quality through our proposed multi-agent framework.

The threats to construct validity mainly lie in the selection of evaluation metrics. To avoid the risk of construct validity, we use the widely used pass@1 metric to evaluate the performance of the code generation models.

\section{Conclusion}
In this paper, we present {\tool} for generating CoT reasoning in multilingual code generation scenarios. 
Our experimental results demonstrate that {\tool} significantly improves the performance of code generation models across different programming languages, achieving competitive results compared to GPT4-based solutions while requiring substantially fewer computational resources.

Future work could explore extending {\tool} to support more programming languages and investigating its application in more complex programming scenarios. 
Additionally, we plan to investigate methods for further improving the efficiency and effectiveness of CoT generation in multilingual environments.

\section*{Acknowledgment}
We thank the anonymous reviewers for their valuable feedback. This research was supported by the National Natural Science Foundation of China under Grant Nos. 62402214 and the Natural Science Foundation of Jiangsu Province under Grant No. BK20241194.

\bibliographystyle{IEEEtran}
\bibliography{reference}

\begin{thebibliography}{10}
\providecommand{\url}[1]{#1}
\csname url@samestyle\endcsname
\providecommand{\newblock}{\relax}
\providecommand{\bibinfo}[2]{#2}
\providecommand{\BIBentrySTDinterwordspacing}{\spaceskip=0pt\relax}
\providecommand{\BIBentryALTinterwordstretchfactor}{4}
\providecommand{\BIBentryALTinterwordspacing}{\spaceskip=\fontdimen2\font plus
\BIBentryALTinterwordstretchfactor\fontdimen3\font minus \fontdimen4\font\relax}
\providecommand{\BIBforeignlanguage}[2]{{%
\expandafter\ifx\csname l@#1\endcsname\relax
\typeout{** WARNING: IEEEtran.bst: No hyphenation pattern has been}%
\typeout{** loaded for the language `#1'. Using the pattern for}%
\typeout{** the default language instead.}%
\else
\language=\csname l@#1\endcsname
\fi
#2}}
\providecommand{\BIBdecl}{\relax}
\BIBdecl

\bibitem{vaithilingam2022expectation}
P.~Vaithilingam, T.~Zhang, and E.~L. Glassman, ``Expectation vs. experience: Evaluating the usability of code generation tools powered by large language models,'' in \emph{Chi conference on human factors in computing systems extended abstracts}, 2022, pp. 1--7.

\bibitem{yang2024important}
G.~Yang, Y.~Zhou, W.~Yang, T.~Yue, X.~Chen, and T.~Chen, ``How important are good method names in neural code generation? a model robustness perspective,'' \emph{ACM Transactions on Software Engineering and Methodology}, vol.~33, no.~3, pp. 1--35, 2024.

\bibitem{wei2022chain}
J.~Wei, X.~Wang, D.~Schuurmans, M.~Bosma, F.~Xia, E.~Chi, Q.~V. Le, D.~Zhou \emph{et~al.}, ``Chain-of-thought prompting elicits reasoning in large language models,'' \emph{Advances in Neural Information Processing Systems}, vol.~35, pp. 24\,824--24\,837, 2022.

\bibitem{jiang2024self}
X.~Jiang, Y.~Dong, L.~Wang, Z.~Fang, Q.~Shang, G.~Li, Z.~Jin, and W.~Jiao, ``Self-planning code generation with large language models,'' \emph{ACM Transactions on Software Engineering and Methodology}, vol.~33, no.~7, pp. 1--30, 2024.

\bibitem{yang2024chain}
G.~Yang, Y.~Zhou, X.~Chen, X.~Zhang, T.~Y. Zhuo, and T.~Chen, ``Chain-of-thought in neural code generation: From and for lightweight language models,'' \emph{IEEE Transactions on Software Engineering}, 2024.

\bibitem{li2023structured}
J.~Li, G.~Li, Y.~Li, and Z.~Jin, ``Structured chain-of-thought prompting for code generation,'' \emph{ACM Transactions on Software Engineering and Methodology}, 2023.

\bibitem{gao2024current}
C.~Gao, X.~Hu, S.~Gao, X.~Xia, and Z.~Jin, ``The current challenges of software engineering in the era of large language models,'' \emph{arXiv preprint arXiv:2412.14554}, 2024.

\bibitem{uchitel2024scoping}
S.~Uchitel, M.~Chechik, M.~Di~Penta, B.~Adams, N.~Aguirre, G.~Bavota, D.~Bianculli, K.~Blincoe, A.~Cavalcanti, Y.~Dittrich \emph{et~al.}, ``Scoping software engineering for ai: The tse perspective,'' \emph{IEEE Transactions on Software Engineering}, vol.~50, no.~11, pp. 2709--2711, 2024.

\bibitem{zheng2023survey}
Z.~Zheng, K.~Ning, Y.~Wang, J.~Zhang, D.~Zheng, M.~Ye, and J.~Chen, ``A survey of large language models for code: Evolution, benchmarking, and future trends,'' \emph{arXiv preprint arXiv:2311.10372}, 2023.

\bibitem{cassano2024knowledge}
F.~Cassano, J.~Gouwar, F.~Lucchetti, C.~Schlesinger, A.~Freeman, C.~J. Anderson, M.~Q. Feldman, M.~Greenberg, A.~Jangda, and A.~Guha, ``Knowledge transfer from high-resource to low-resource programming languages for code llms,'' \emph{Proceedings of the ACM on Programming Languages}, vol.~8, no. OOPSLA2, pp. 677--708, 2024.

\bibitem{yang2024multi}
H.~Yang, Y.~Nong, S.~Wang, and H.~Cai, ``Multi-language software development: Issues, challenges, and solutions,'' \emph{IEEE Transactions on Software Engineering}, 2024.

\bibitem{peng2024humaneval}
Q.~Peng, Y.~Chai, and X.~Li, ``Humaneval-xl: A multilingual code generation benchmark for cross-lingual natural language generalization,'' in \emph{Proceedings of the 2024 Joint International Conference on Computational Linguistics, Language Resources and Evaluation (LREC-COLING 2024)}, 2024, pp. 8383--8394.

\bibitem{cohn2010inducing}
T.~Cohn, P.~Blunsom, and S.~Goldwater, ``Inducing tree-substitution grammars,'' \emph{The Journal of Machine Learning Research}, vol.~11, pp. 3053--3096, 2010.

\bibitem{allamanis2014mining}
M.~Allamanis and C.~Sutton, ``Mining idioms from source code,'' in \emph{Proceedings of the 22nd acm sigsoft international symposium on foundations of software engineering}, 2014, pp. 472--483.

\bibitem{gulwani2010dimensions}
S.~Gulwani, ``Dimensions in program synthesis,'' in \emph{Proceedings of the 12th international ACM SIGPLAN symposium on Principles and practice of declarative programming}, 2010, pp. 13--24.

\bibitem{zan2023large}
D.~Zan, B.~Chen, F.~Zhang, D.~Lu, B.~Wu, B.~Guan, W.~Yongji, and J.-G. Lou, ``Large language models meet nl2code: A survey,'' in \emph{Proceedings of the 61st Annual Meeting of the Association for Computational Linguistics (Volume 1: Long Papers)}, 2023, pp. 7443--7464.

\bibitem{liu2016automatic}
Z.~Liu, Y.~Dou, J.~Jiang, and J.~Xu, ``Automatic code generation of convolutional neural networks in fpga implementation,'' in \emph{2016 International conference on field-programmable technology (FPT)}.\hskip 1em plus 0.5em minus 0.4em\relax IEEE, 2016, pp. 61--68.

\bibitem{sun2019grammar}
Z.~Sun, Q.~Zhu, L.~Mou, Y.~Xiong, G.~Li, and L.~Zhang, ``A grammar-based structural cnn decoder for code generation,'' in \emph{Proceedings of the AAAI conference on artificial intelligence}, vol.~33, no.~01, 2019, pp. 7055--7062.

\bibitem{iyer2016summarizing}
S.~Iyer, I.~Konstas, A.~Cheung, and L.~Zettlemoyer, ``Summarizing source code using a neural attention model,'' in \emph{Proceedings of the 54th Annual Meeting of the Association for Computational Linguistics (Volume 1: Long Papers)}.\hskip 1em plus 0.5em minus 0.4em\relax Association for Computational Linguistics, 2016.

\bibitem{wan2018improving}
Y.~Wan, Z.~Zhao, M.~Yang, G.~Xu, H.~Ying, J.~Wu, and S.~Y. Philip, ``Improving automatic source code summarization via deep reinforcement learning,'' in \emph{2018 33rd IEEE/ACM International Conference on Automated Software Engineering (ASE)}.\hskip 1em plus 0.5em minus 0.4em\relax IEEE Computer Society, 2018, pp. 397--407.

\bibitem{yin2017syntactic}
P.~Yin and G.~Neubig, ``A syntactic neural model for general-purpose code generation,'' in \emph{Proceedings of the 55th Annual Meeting of the Association for Computational Linguistics (Volume 1: Long Papers)}, 2017, pp. 440--450.

\bibitem{vaswani2017attention}
A.~Vaswani, ``Attention is all you need,'' \emph{Advances in Neural Information Processing Systems}, 2017.

\bibitem{mastropaolo2021studying}
A.~Mastropaolo, S.~Scalabrino, N.~Cooper, D.~N. Palacio, D.~Poshyvanyk, R.~Oliveto, and G.~Bavota, ``Studying the usage of text-to-text transfer transformer to support code-related tasks,'' in \emph{2021 IEEE/ACM 43rd International Conference on Software Engineering (ICSE)}.\hskip 1em plus 0.5em minus 0.4em\relax IEEE, 2021, pp. 336--347.

\bibitem{shah2021natural}
M.~Shah, R.~Shenoy, and R.~Shankarmani, ``Natural language to python source code using transformers,'' in \emph{2021 International Conference on Intelligent Technologies (CONIT)}.\hskip 1em plus 0.5em minus 0.4em\relax IEEE, 2021, pp. 1--4.

\bibitem{feng2020codebert}
Z.~Feng, D.~Guo, D.~Tang, N.~Duan, X.~Feng, M.~Gong, L.~Shou, B.~Qin, T.~Liu, D.~Jiang \emph{et~al.}, ``Codebert: A pre-trained model for programming and natural languages,'' in \emph{Findings of the Association for Computational Linguistics: EMNLP 2020}, 2020, pp. 1536--1547.

\bibitem{lu1codexglue}
S.~Lu, D.~Guo, S.~Ren, J.~Huang, A.~Svyatkovskiy, A.~Blanco, C.~Clement, D.~Drain, D.~Jiang, D.~Tang \emph{et~al.}, ``Codexglue: A machine learning benchmark dataset for code understanding and generation,'' in \emph{Thirty-fifth Conference on Neural Information Processing Systems Datasets and Benchmarks Track (Round 1)}.

\bibitem{ahmad2021unified}
W.~Ahmad, S.~Chakraborty, B.~Ray, and K.-W. Chang, ``Unified pre-training for program understanding and generation,'' in \emph{Proceedings of the 2021 Conference of the North American Chapter of the Association for Computational Linguistics: Human Language Technologies}, 2021, pp. 2655--2668.

\bibitem{wang2021codet5}
Y.~Wang, W.~Wang, S.~Joty, and S.~C. Hoi, ``Codet5: Identifier-aware unified pre-trained encoder-decoder models for code understanding and generation,'' in \emph{Proceedings of the 2021 Conference on Empirical Methods in Natural Language Processing}, 2021, pp. 8696--8708.

\bibitem{chen2021evaluating}
M.~Chen, J.~Tworek, H.~Jun, Q.~Yuan, H.~P. D.~O. Pinto, J.~Kaplan, H.~Edwards, Y.~Burda, N.~Joseph, G.~Brockman \emph{et~al.}, ``Evaluating large language models trained on code,'' \emph{arXiv preprint arXiv:2107.03374}, 2021.

\bibitem{li2022competition}
Y.~Li, D.~Choi, J.~Chung, N.~Kushman, J.~Schrittwieser, R.~Leblond, T.~Eccles, J.~Keeling, F.~Gimeno, A.~Dal~Lago \emph{et~al.}, ``Competition-level code generation with alphacode,'' \emph{Science}, vol. 378, no. 6624, pp. 1092--1097, 2022.

\bibitem{friedincoder}
D.~Fried, A.~Aghajanyan, J.~Lin, S.~Wang, E.~Wallace, F.~Shi, R.~Zhong, S.~Yih, L.~Zettlemoyer, and M.~Lewis, ``Incoder: A generative model for code infilling and synthesis,'' in \emph{The Eleventh International Conference on Learning Representations}.

\bibitem{nijkampcodegen}
E.~Nijkamp, B.~Pang, H.~Hayashi, L.~Tu, H.~Wang, Y.~Zhou, S.~Savarese, and C.~Xiong, ``Codegen: An open large language model for code with multi-turn program synthesis,'' in \emph{The Eleventh International Conference on Learning Representations}.

\bibitem{nijkampcodegen2}
E.~Nijkamp, H.~Hayashi, C.~Xiong, S.~Savarese, and Y.~Zhou, ``Codegen2: Lessons for training llms on pro-gramming and natural languages.''

\bibitem{li2023starcoder}
R.~Li, L.~Allal, Y.~Zi, N.~Muennighoff, D.~Kocetkov, C.~Mou, M.~Marone, C.~Akiki, J.~Li, J.~Chim \emph{et~al.}, ``Starcoder: May the source be with you!'' \emph{Transactions on machine learning research}, 2023.

\bibitem{luowizardcoder}
Z.~Luo, C.~Xu, P.~Zhao, Q.~Sun, X.~Geng, W.~Hu, C.~Tao, J.~Ma, Q.~Lin, and D.~Jiang, ``Wizardcoder: Empowering code large language models with evol-instruct,'' in \emph{The Twelfth International Conference on Learning Representations}.

\bibitem{muennighoff2023octopack}
N.~Muennighoff, Q.~Liu, A.~Zebaze, Q.~Zheng, B.~Hui, T.~Y. Zhuo, S.~Singh, X.~Tang, L.~Von~Werra, and S.~Longpre, ``Octopack: Instruction tuning code large language models,'' in \emph{NeurIPS 2023 Workshop on Instruction Tuning and Instruction Following}.

\bibitem{roziere2023code}
B.~Rozi{\`e}re, J.~Gehring, F.~Gloeckle, S.~Sootla, I.~Gat, X.~E. Tan, Y.~Adi, J.~Liu, T.~Remez, J.~Rapin \emph{et~al.}, ``Code llama: Open foundation models for code,'' \emph{arXiv e-prints}, pp. arXiv--2308, 2023.

\bibitem{wei2022emergent}
J.~Wei, Y.~Tay, R.~Bommasani, C.~Raffel, B.~Zoph, S.~Borgeaud, D.~Yogatama, M.~Bosma, D.~Zhou, D.~Metzler \emph{et~al.}, ``Emergent abilities of large language models,'' \emph{Transactions on Machine Learning Research}, 2022.

\bibitem{zhang2024self}
X.~Zhang, B.~Peng, Y.~Tian, J.~Zhou, L.~Jin, L.~Song, H.~Mi, and H.~Meng, ``Self-alignment for factuality: Mitigating hallucinations in llms via self-evaluation,'' \emph{arXiv preprint arXiv:2402.09267}, 2024.

\bibitem{yang2024less}
G.~Yang, Y.~Zhou, X.~Zhang, W.~Cheng, K.~Liu, X.~Chen, T.~Y. Zhuo, and T.~Chen, ``Less is more: Towards green code large language models via unified structural pruning,'' \emph{arXiv preprint arXiv:2412.15921}, 2024.

\bibitem{wei2024magicoder}
Y.~Wei, Z.~Wang, J.~Liu, Y.~Ding, and L.~Zhang, ``Magicoder: Empowering code generation with oss-instruct,'' in \emph{Forty-first International Conference on Machine Learning}, 2024.

\bibitem{guo2024deepseek}
D.~Guo, Q.~Zhu, D.~Yang, Z.~Xie, K.~Dong, W.~Zhang, G.~Chen, X.~Bi, Y.~Li \emph{et~al.}, ``Deepseek-coder: When the large language model meets programming--the rise of code intelligence,'' \emph{arXiv preprint arXiv:2401.14196}, 2024.

\bibitem{hui2024qwen2}
B.~Hui, J.~Yang, Z.~Cui, J.~Yang, D.~Liu, L.~Zhang, T.~Liu, J.~Zhang, B.~Yu, K.~Lu \emph{et~al.}, ``Qwen2. 5-coder technical report,'' \emph{arXiv preprint arXiv:2409.12186}, 2024.

\end{thebibliography}

\end{document}